\documentclass[aip,amsmath,amssymb,reprint]{revtex4-1}

\usepackage{graphicx}
\usepackage{dcolumn}
\usepackage{bm}
\usepackage[utf8]{inputenc}
\usepackage[T1]{fontenc}
\usepackage{mathptmx}
\usepackage{etoolbox}
\newcommand{\numsub}[1]{$_{#1}$}

\makeatletter
\def\@email#1#2{%
 \endgroup
 \patchcmd{\titleblock@produce}
  {\frontmatter@RRAPformat}
  {\frontmatter@RRAPformat{\produce@RRAP{*#1\href{mailto:#2}{#2}}}\frontmatter@RRAPformat}
  {}{}
}%
\makeatother
\begin{document}

\title[Oxide nanoscrolls for H$_2$ generation and storage]{Robust and tunable oxide nanoscrolls for solar-driven H$_2$ generation and storage}
\author{Adway Gupta}
\author{Arunima K. Singh}
 \email{arunimasingh@asu.edu}
\affiliation{Department of Physics, Arizona State University, Tempe, Arizona  85287-1504}
\date{\today}

\begin{abstract}

Hydrogen gas is a promising alternative to fossil fuels due to its high energy output and environmentally safe byproducts. Various morphologies of photocatalytic materials have been explored for high-efficiency H$_2$ production, for instance, quasi-1D nanoscroll structures that provide larger surface-to-volume ratio. Recently, we predicted layer-by-layer formation of stable oxide nanoscrolls directly from dichalcogenide precursors, eliminating the need for costly formation of two-dimensional oxides for a roll-up synthesis of nanoscrolls. In this study, we evaluate the suitability of those oxide nanoscroll materials— MoO\numsub{3}, WO\numsub{3}, PdO\numsub{2}, HfO\numsub{2}, and GeO\numsub{2}—for solar-driven photocatalytic H\numsub{2} production and storage. Using excited state theory simulations we discern their electronic properties as a function of interlayer scroll spacing and find them to possess electronic properties that are suitable for photocatalysis. Additionally, using \textit{ab initio} molecular dynamics simulations we show that they are also suitable for H\numsub{2} storage as the nanoscrolls exhibit effective trapping of hydrogen, even in the presence of defects and vacancies in the oxides. This work thus demonstrates the discovery of robust and tunable oxide nanoscrolls as novel materials for advancing solar-driven hydrogen technologies.
\end{abstract}

\maketitle

Molecular hydrogen (H\numsub{2}) is a promising alternative to traditional fossil fuels due to its high energy output and harmless byproducts\cite{kumar2022overview,song2022solar}. Beyond its potential as a clean energy source, hydrogen also plays a crucial role as a raw material in industries such as those for methane production and atmospheric CO\numsub{2} capture, further enhancing its value in the transition to a sustainable energy future\cite{tee2017recent,holladay2009overview}.

Current large-scale hydrogen production relies heavily on fossil fuels, undermining its potential to be a renewable energy source. However, hydrogen can also be produced by splitting earth-abundant water. The Gibbs free energy of water splitting is 1.23 V, thus at least 1.23 eV of energy is required to produce H$_2$ from water. This energy can be obtained by a solar-driven water-splitting using a suitable photocatalyst. Photocatalytic water splitting was first demonstrated by Honda and Fujishima's\cite{fujishima1972electrochemical}, in  TiO\numsub{2}, and since then several metal oxide photocatalysts like BiVO\numsub{4} and WO\numsub{3} have been discovered\cite{h2_storage_tmo1,h2_storage_tmo2,costa2022current,hisatomi2014advances,ahmad_review}.
While these oxides are excellent photocatalysts, their efficiencies can be further enhanced by increasing the reaction surface area and solving the problem of long carrier travel distances that increase recombination rates and reduce efficiency. Nanostructuring, such as formation of two-dimensional (2D) films, nanotubes, or nanoscrolls, can shorten the charge carrier distances and increase the surface-to-volume ratio, enhancing both water-splitting efficiency and H\numsub{2} adsorption. 




\begin{figure}[ht]
    \centering
    \includegraphics[width=3.25in]{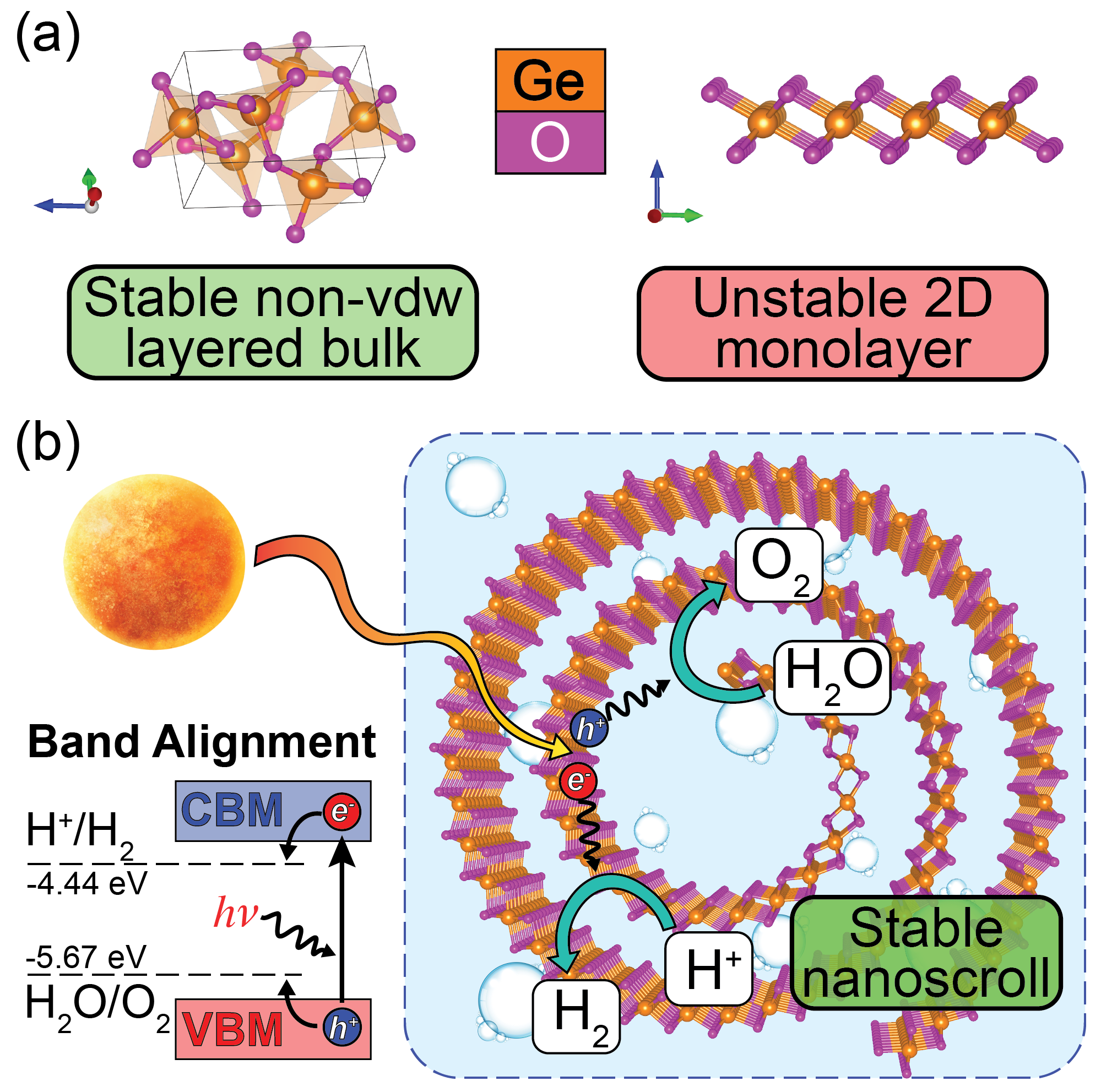}
    \caption{(a) Shows the crystal structure of bulk GeO\numsub{2}. While GeO\numsub{2} has a stable non-vdW layered bulk structure, it has no known stable 2D counterpart. (b) Shows a schematic for a stable GeO$_2$ nanoscroll functioning as a solar-driven water-splitting photocatalyst. Band edges required for the hydrogen evolution reaction(HER) and oxygen evolution reaction(OER) half reactions are shown.}
    \label{introduction_fig}
\end{figure}
 
Among nanostructures, quasi-1D nanoscroll structures offer large surface-to-volume ratios and shorter carrier distances, and their closed morphology allows H\numsub{2} to be stored both on the surface and between the layers\cite{lee2021photoelectrochemical,khan2021quasi,daulbayev20200d}. Additionally, the physical parameters of nanoscrolls—such as inner radius, core radius, and interlayer spacing—can be easily tuned, enabling precise control over their electronic and transport properties for optimized solar-driven H\numsub{2} production and storage. 

Recently, we developed an $ab$ $initio$ physio-mechanical model to predict stable nanoscroll structures and discovered stable MoO\numsub{3}, WO\numsub{3}, PdO\numsub{2}, HfO\numsub{2} and 1T-GeO\numsub{2} scrolls. The crystal structures of the oxide materials are shown in supplementary figure S1. 



In this manuscript, we examine the suitability of these oxide nanoscrolls for solar-driven H\numsub{2} production and H\numsub{2} storage via $ab initio$ simulations. Using van der Waals(vdW) corrected excited state theory $G$\numsub{0}$W$\numsub{0} simulations we compute their band structures and band edges relative to vacuum and also study their dependence on the interlayer spacing of the nanoscrolls. To demonstrate the application of these materials as H\numsub{2} storage materials, we additionally perform \textit{ab initio} molecular dynamics (AIMD) simulations of hydrogen species – H, H$^+$ and H\numsub{2}, intercalated between layers of the scroll. Finally, we also model the transport of these hydrogen species in the presence of cation vacancies in the oxide structures under an applied electric field. The results obtained in this work will guide the experimental realization of robust oxide nanoscroll structures for solar-driven H\numsub{2} production and storage. 

Given that the primary interaction between the scroll layers are vdW interactions, in order to model the dependence of the electronic structure of the nanoscrolls on the interlayer spacing, we use the widely used approximation of flat and infinite 2D nanosheets of the oxide material, with the interlayer spacing $h$ of the nanoscroll represented by the periodic separation in the $z$-direction, $h$, between adjacent 2D monolayers, as shown for MoO\numsub{3} in Supplementary figure S15\cite{nsapprox1,nsapprox2,nsapprox3,nsapprox4,nsapprox5,nsapprox6}. To establish a lower limit for the interlayer spacing below which the nanoscroll structure would be unstable, we use vdW-corrected Density Functional Theory (DFT) calculations to relax the unit cell of the oxide monolayers along the $z$-direction such that forces on the atoms are minimized. The interlayer spacings in this resultant structure is the lower limit of the interlayer spacings, $\Delta z_{\text{exact}}^{\text{min}}$. 
To study the effect of interlayer spacing, we begin with 2D nanosheets at this minimum separation, $\Delta z_{\text{exact}}^{\text{min}}$, and increase the spacing by 5 \AA, 10 \AA, and 15 \AA. Where $\Delta z_{\text{exact}}$ is the distance between the top of one layer to the bottom of its periodic image in the $z$-direction, as shown in Supplementary figure S2. To eliminate all long-range vdW interactions, we also simulate a final configuration with $\Delta z_{\text{exact}} > 20 \AA$, resulting in a total of five configurations for each oxide.


\begin{figure}
    \centering
    \includegraphics[width=\linewidth]{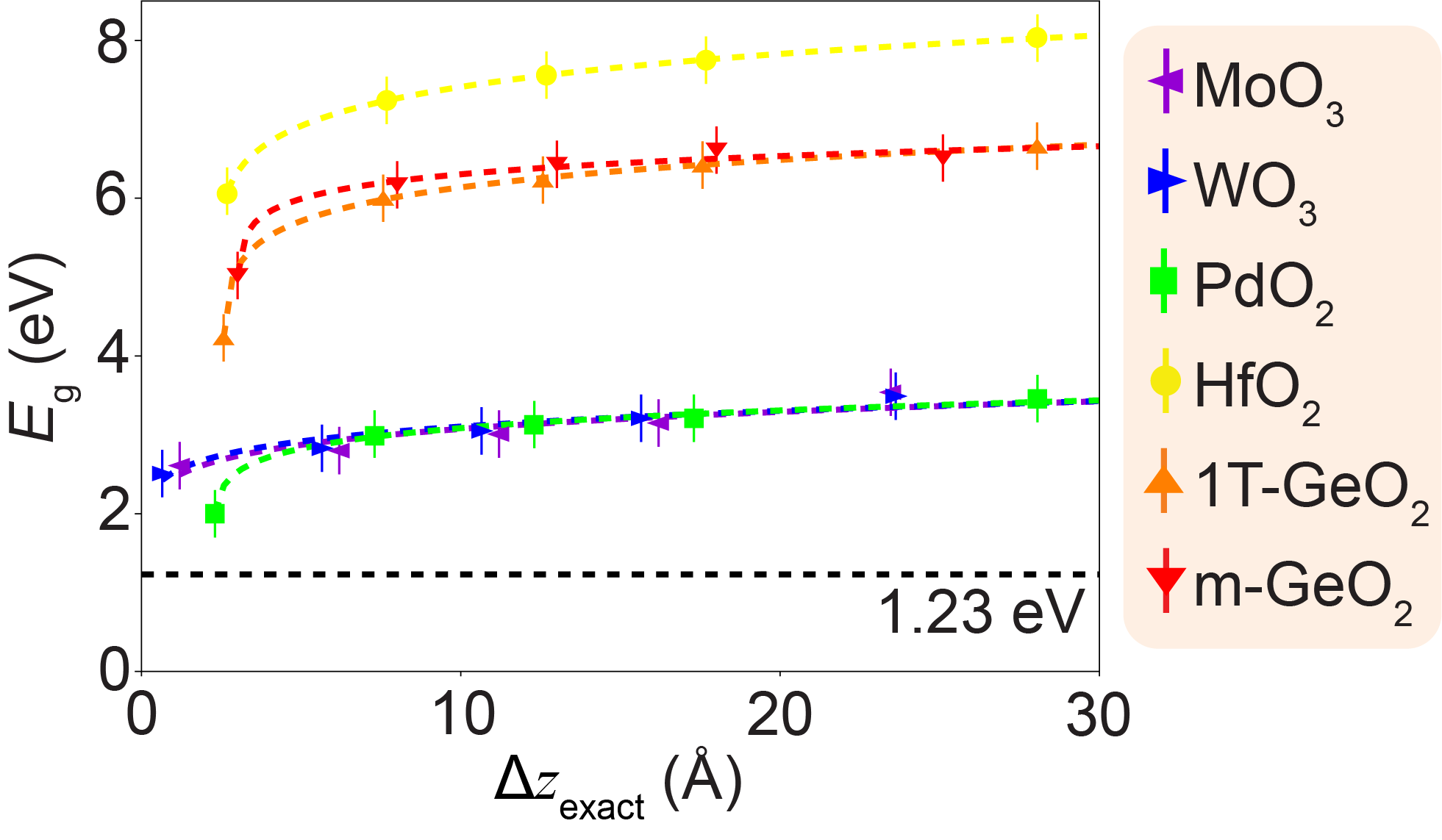}
    \caption{Figure shows the evolution of the $GW$ computed bandgaps with interlayer scroll spacing for all our selected structures. The convergence in the bandgaps is 0.15 eV and is marked by the error bars. It can be clearly seen from the figure that all the bandgaps lie above the threshold of 1.23 eV required for solar-driven H\numsub{2} production}
    \label{band_gaps_fig}
\end{figure}

\begin{figure*}[ht]
    \centering
    \includegraphics[width=\linewidth]{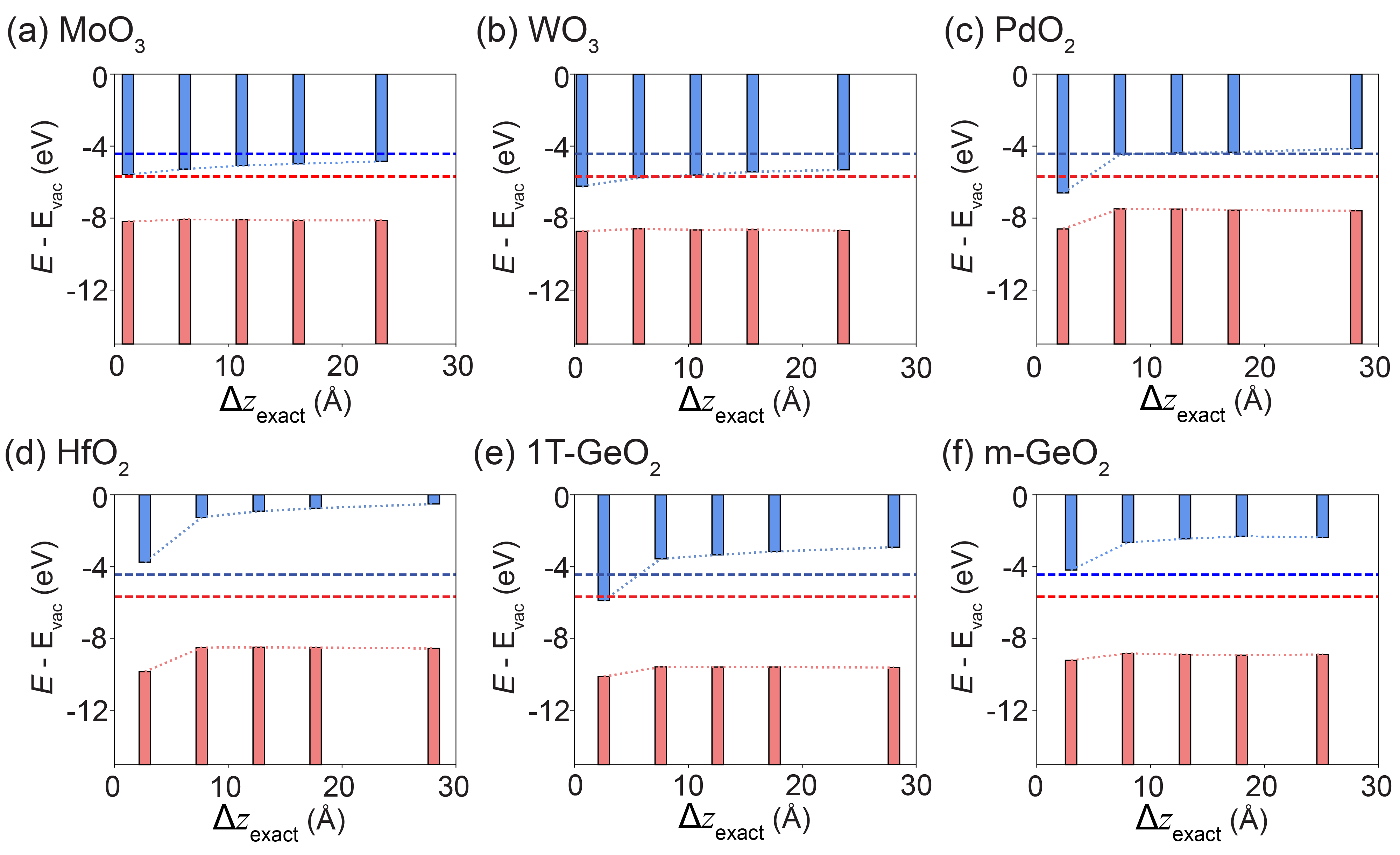}
    \caption{Figure shows the evolution of band-edges with interlayer spacing for all our selected materials. Blue and red dashed lines show the potentials for the HER and OER respectively. It can be seen that the valence band edge lies below the OER potential for all the materials.}
    \label{band_edges_fig}
\end{figure*}

To accurately determine the electronic structure of the described configurations, we use vdW-corrected excited state many-body perturbation theory G\numsub{0}W\numsub{0} simulations, instead of DFT since DFT is known to grossly underestimate bandgaps. All the DFT simulations for determining the trial wavefunctions for the G\numsub{0}W\numsub{0} step, are based on the projector-augmented wave method as implemented in the plane-wave code VASP\cite{Kresse1,Kresse2,Kresse3,Kresse4,Kresse5}. To account for the vdW interactions, simulations were performed using the vdW-DF-optB88 exchange–correlation functional\cite{Klimes2011}, with a cutoff energy for the plane wave basis at 600 eV and a $k$-grid density of 50 \AA$^{-3}$. Structural relaxations for the oxide monolayers were performed with energy convergence within $10^{-6}$ eV for each ionic step and until the forces were converged to $0.05$ eV/atom. The quasiparticle (QP) energies were obtained using many-body perturbation theory within the G\numsub{0}W\numsub{0} approximation for the self-energy operator. Other parameters for the G\numsub{0}W\numsub{0} calculations can be found in the supplementary table S3. 
To obtain the QP band structure, we used the interpolation formalism implemented in the WANNIER90 package\cite{Pizzi2020}.

Figure \ref{band_gaps_fig} shows the bandgaps of all selected oxides across the five interlayer spacings. The data points are fitted with trend lines, and error bars are included for clarity. The values of the bandgaps and if the bandgap is direct or not is available in supplementary table S2. All the bandstructures for each interlayer spacing and oxide are shown in supplementary figures S3-S8. We can clearly see from Fig \ref{band_gaps_fig} that the bandgaps for all the oxides across the entire range of interlayer spacings are above the 1.23 eV threshold that is required for water-splitting, indicating all the selected materials can potentially generate the appropriate charge carriers for the HER and OER half-reactions.
Furthermore, we see that the bandgaps of MoO\numsub{3}, WO\numsub{3} and PdO\numsub{2} are less than the threshold for absorption in the visible range, which is around 3 eV making them ideal photocatalysts in the visible range.
On the other hand, HfO\numsub{2}, 1T-GeO\numsub{2} and m-GeO\numsub{2} can find use in the near ultraviolet spectrum. 

In general, we find that the bandgap increases with interlayer spacing as a result of the quantum confinement, which is consistent with the trends from other 2D materials like MoS\numsub{2}. Finally, we see that as expected, the band gaps converge to a constant value with increasing values of the interlayer separation.


Fig \ref{band_edges_fig} shows the variation of the band-edges with respect to the interlayer spacing of the nanoscrolls for all our selected oxides. To compute the band-edges with respect to vacuum, we first performed the electrostatic potential calculations at the DFT level, then used the band gap center (BGC) approximation to obtain the \textit{GW} computed band edges\cite{bgcapprox1,bgcapprox2}. In the BCG approximation,
\begin{equation}\label{bgc_approximation}
    E_{\text{CBM/VBM}}^{\text{vac}}(\textit{GW}) = E_{\text{BGC}}(\textit{GW}) - E^{\text{vac}}(\text{DFT}) \pm E_{\text{g}}(\textit{GW})
\end{equation}
Where $E_{\text{g}}(\text{GW})$ is the G\numsub{0}W\numsub{0} computed bandgap of the configuration, $E^{\text{vac}}(\text{DFT})$ is the DFT computed vacuum level and $E_{\text{BGC}}(\textit{GW}) =(E_{\text{CBM}}(\textit{GW}) + E_{\text{VBM}}(\textit{GW}))/2$. Further details about the determination of the band edges are given in the supplementary information. \\

In fig \ref{band_edges_fig} the potential of the HER at -4.44 eV and OER at -5.67 eV is shown with blue and red dashed lines respectively. We can clearly see that for all the structures the VBM edge lies lower than the potential for the OER, which is -5.67 eV, implying that all of these nanoscrolls can readily act as photoanodes and supply holes from their valence band in order to promote oxygen evolution, across their entire range of interlayer spacings.
Furthermore, the $E_{\text{CBM}}^{\text{vac}}(\textit{GW})$ of HfO\numsub{2} and m-GeO\numsub{2} nanoscrolls are above the -4.44 eV for the HER reaction, indicating that they can act as single photocatalysts across their entire range of interlayer spacings.
For 1T-GeO\numsub{2} and PdO\numsub{2}, we can see their effectiveness as single photocatalysts is limited to specific ranges of interlayer spacings.
Finally, for MoO\numsub{3} and WO\numsub{3}, we observe that their $E_{\text{CBM}}^{\text{vac}}(\textit{GW})$ remains below the HER potential across all interlayer spacings. While this indicates that spontaneous hydrogen evolution is not feasible with these two TMO nanoscrolls, it is possible to utilize them for HER By applying an overpotential or selecting an appropriate cocatalyst to supply electrons for the reduction of H$^+$ ions\cite{cocatalyst2,cocatalysts1}. Thus, we find that the electronic structures of our selected oxide nanoscrolls are highly tunable by varying the interlayer spacing, allowing controllable solar-driven photocatalytic applications. 


\begin{figure*}
    \centering
    \includegraphics{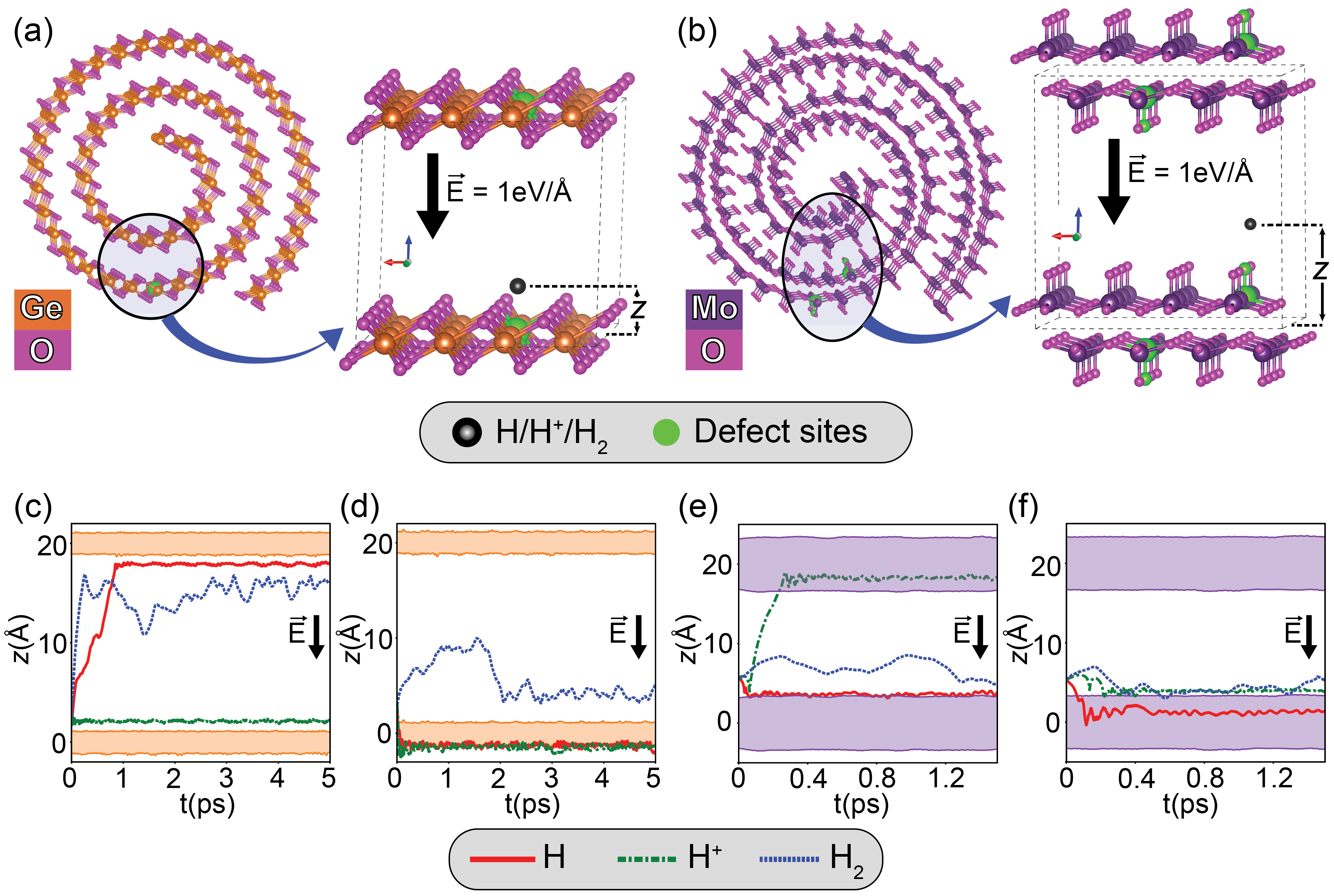}
    \caption{(a) and (b) show the schematic modeling of the interlayer transport of H, H$^+$ and H\numsub{2} in the presence of an electric field for 1T-GeO\numsub{2} and MoO\numsub{3} nanoscrolls respectively. Atoms in green show the location of the cation defect. (c) and (d) show the time evolution of the $z$-position for H (solid red), H$^+$ (dot-dash green) and H\numsub{2} (dotted blue) for pristine and defected 1T-GeO\numsub{2} respectively. (e) and (f) show the same information for pristine and defected MoO\numsub{3} respectively.}
    \label{MDfig}
\end{figure*}

As previously mentioned, nanoscroll structures show great promise for hydrogen storage due to their unique morphologies,  allowing for the confinement of H\numsub{2} molecules in the regions between the nanoscroll layers and the center, in addition to adsorption on their surfaces. For our selected nanoscroll to act as simultaneous H\numsub{2} generation and H\numsub{2} storage materials, we require the effective adsorption of H atoms or H$^+$ ions on the nanoscroll surface, enabling efficient electron transfer for the HER. Furthermore, for the resulting H\numsub{2} molecules to be effectively utilized, they must remain trapped between the scroll layers. Thus, we evaluated the transport of three intercalated hydrogen species – H atom, H$^+$ ion, and H\numsub{2} molecule between the layers of the scroll. We performed this analysis in the presence of a 1 eV/\AA transverse electric field that can provide a driving force to mobilize the intercalated species. We examined the effect of this molecular transport on the species as well as the structure of the oxides. Since these simulations are performed for the computationally expensive AIMD simulations, we performed this study of adsorption and transport of species on only two representative nanoscrolls- 1T-GeO\numsub{2} and MoO\numsub{3}.

 All the AIMD simulations were done using the plane-wave code VASP\cite{Kresse1,Kresse2,Kresse3,Kresse4,Kresse5}. We simulated 4x4x1 supercells of 1T-GeO\numsub{2} and MoO\numsub{3} as shown in Fig \ref{MDfig}(a) and (b) respectively. The intercalated species were placed at an initial height of 2 \AA above the materials. A transverse field of 1 eV/\AA was applied to study the molecular transport. we modeled the nanoscrolls with 2D analogs with an exaggerated interlayer spacing of 20 \AA in order to clearly simulate the transport properties. Based on its electronic properties, at this interlayer spacing, 1T-GeO\numsub{2} can act as a single photocatalyst. The AIMD simulations were performed in the NVT ensemble with a Langevin thermostat\cite{allen2017computer} at 300 K, with a 40x40x1 $k$-grid. A timestep of 1 fs was selected and the simulations were run for 5000 and 1500 timesteps, corresponding to total times of 5 ps and 1.5 ps for 1T-GeO\numsub{2} and MoO\numsub{3}, respectively. Within each timestep, the total energies were converged within $10^{-5}$ eV. The $z$-coordinates of the intercalated species were referenced to the oxide layers over time. 
 
 Fig \ref{MDfig} (c) and (e) show the $z$-coordinates of the three species, H atom, H$^+$ ion, and H$_2$ molecule as a function of time under the applied field for 1T-GeO\numsub{2} and MoO\numsub{3}, respectively. The $z$-coordinates for the H atom, H$^+$ ion and H\numsub{2} molecule are plotted with solid red lines, dotted-dashed green lines, and dashed blue lines respectively. Additionally, the shaded regions in the figures show the region occupied by the 2D sheets in the $z$-direction over time. 
 
 From Figs \ref{MDfig} (c) and (e) we can see that both H atoms and H$^+$ ions are effectively adsorbed on the surfaces of the oxide nanoscrolls, in both 1T-GeO\numsub{2} and MoO\numsub{3}. This adsorption suggests that the nanoscrolls offer sufficient active sites for the HER to occur, facilitating the formation of H\numsub{2}. On the other hand, H\numsub{2} molecules remain trapped between the layers of the scroll, indicating that the scroll morphology allows for effective trapping and storage of molecular H\numsub{2}. Furthermore, the nanoscroll surfaces remain intact throughout the period of the simulation and do not disintegrate or distort upon impact by the intercalated species
 Hence, we can surmise that the selected nanoscrolls can perform effectively as simultaneous H\numsub{2} production and storage materials at room temperature. 

\begin{figure}
    \centering
    \includegraphics[width=3.33 in]{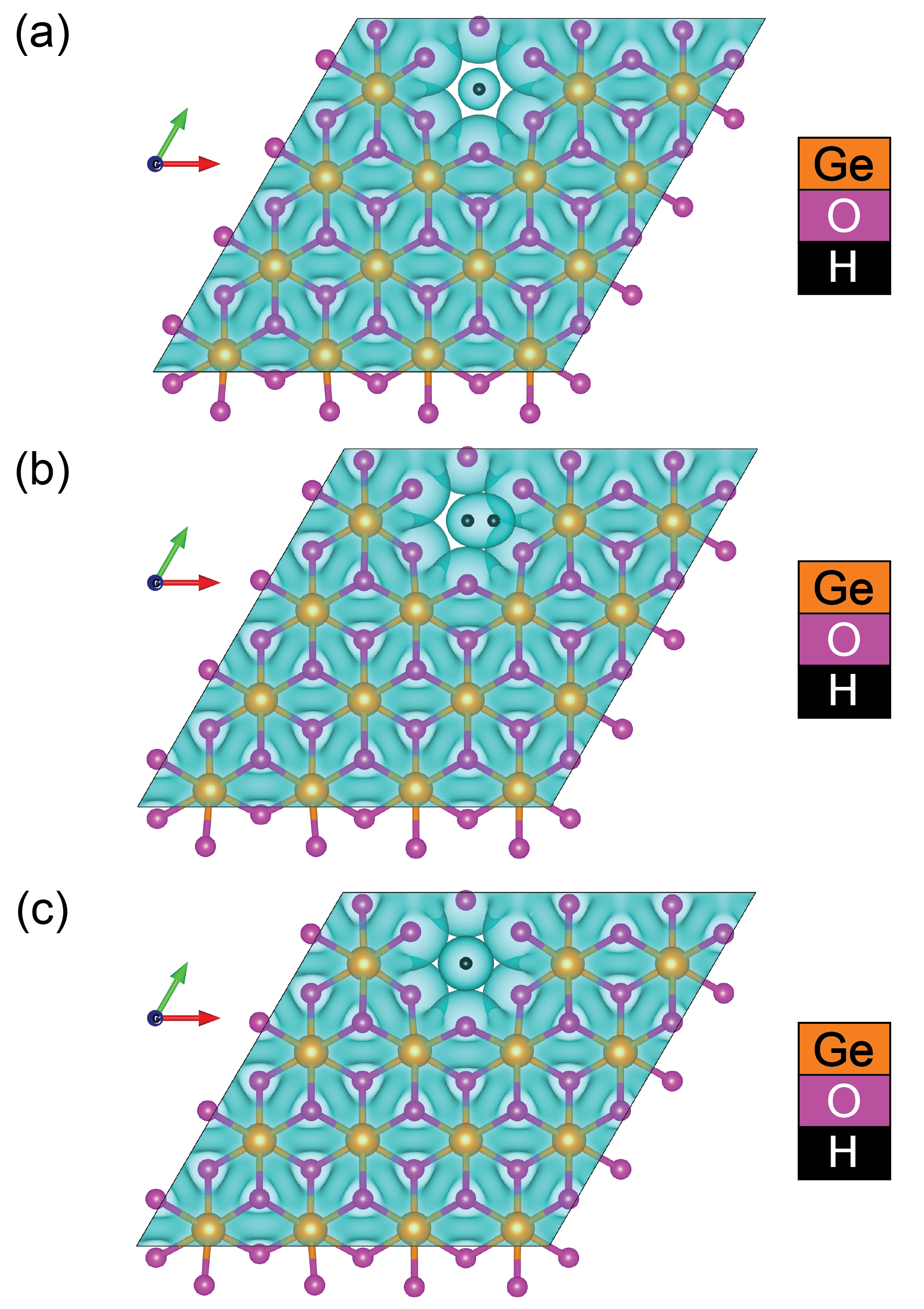}
    \caption{Figures show the $z$-projection of the charge densities of an (a) H atom, (b) H\numsub{2} molecule placed transverse to the $z$-direction and (c) H\numsub{2} molecule placed parallel to the $z$-direction, located above a Ge defect in a 4x4x1 1T-GeO\numsub{2} supercell. All charge densities are plotted with an isosurface level of 0.03 $e^-$/\AA$^3$ }
    \label{charge_densities_figure}
\end{figure}

However, in practical applications, defects in materials can impact their interactions with intercalated species. We introduce metal vacancies in the simulation cell to examine the effect of cation vacancies on the transport and storage of hydrogen species. As shown in Fig \ref{MDfig} (a) and (b), we created one Ge, and 2 Mo vacancies in the 4x4x1 1T-GeO\numsub{2} and MoO\numsub{3} supercells, respectively. These correspond to planar defect densities of $8.44\times 10^{-3}$\AA$^{-2}$ and $8.62\times 10^{-3}$\AA$^{-2}$ for 1T-GeO\numsub{2} and MoO\numsub{3}, respectively. The intercalated hydrogen species were again placed 2 \AA directly above the defect site. The remaining simulation parameters were the same as the ones for the pristine case. We tracked the $z$-coordinate of the intercalated species over time and the results for 1T-GeO\numsub{2} and MoO\numsub{3} are shown in Fig \ref{MDfig} (d) and (f), respectively. 

We can see that in the presence of the vacancies, the oxide materials retained their morphology without disintegrating, establishing their robustness. We also see that in the presence of the transverse electric field, H atoms and H$^+$ ions were highly susceptible to passing through the cation defect site in both 1T-GeO\numsub{2} and MoO\numsub{3}. In 1T-GeO\numsub{2}, after traversing this defect, H and H$^+$ were found to be adsorbed on the opposite side of the layer. In contrast, in MoO\numsub{3}, the H atom passed through the surface and become trapped within the layer, most likely due to the larger thickness of MoO\numsub{3} in comparison to that of 1T-GeO\numsub{2}. In MoO\numsub{3}, the H$^+$ ion remained adsorbed on the surface. Thus, both H and H$^+$ ions remained bound to the surfaces, similar to what was observed in pristine structures, and in turn provide the active sites necessary for the HER.

Figure \ref{MDfig}(d) and (f), show that, surprisingly, the cation vacancy plays no role in the opacity of the oxide layers to the H\numsub{2} molecule. The H\numsub{2} molecule freely moves in the region between the layers and does not tunnel through the defect site. As a result, the defected materials maintain their propensity for H\numsub{2} storage. 

To understand the relative permeability of H atoms and H$^+$ ions compared to H\numsub{2} molecules, in Fig \ref{charge_densities_figure} we plot the charge densities of the pristine and defected 1T-GeO\numsub{2} structure. Fig \ref{charge_densities_figure}(a) shows the charge density of a single hydrogen atom placed directly above the defect site in a 4x4x1 1T-GeO\numsub{2} structure. Fig \ref{charge_densities_figure} (b) and (c) show the charge densities of an H\numsub{2} molecule placed transverse and parallel to the $z$-direction respectively, again directly above the defect site. All charge densities are plotted at an isosurface level of 0.03 $e^-$/\AA$^3$. We can clearly see from Fig \ref{charge_densities_figure}(a) that the charge density of the H atom is smaller than the vacancy site. This allows it to travel through the material.

On the other hand, Fig \ref{charge_densities_figure}(b)-(c) clearly shows that the charge density of the H\numsub{2} molecule is larger than the vacancy site. As a result, it experiences large Coulomb repulsion from the 1T-GeO\numsub{2} structure and it cannot pass through it,  remaining confined within its layers.

In summary, in this study, we studied quasi-1D nanoscroll structures of MoO\numsub{3}, WO\numsub{3}, PdO\numsub{2}, HfO\numsub{2}, 1T-GeO\numsub{2}, and m-GeO\numsub{2} to establish their suitability for photocatalytic water-splitting and hydrogen intercalation. We determined their band gaps across various interlayer spacings, using the G$_0$W$_0$ approximation. We find that the nanoscrolls possess band gaps exceeding the 1.23 eV threshold required for photocatalytic water-splitting. Notably, MoO\numsub{3}, WO\numsub{3}, and PdO\numsub{2} exhibit band gaps within the visible spectrum, positioning them as ideal candidates for solar-driven applications. We also examined the band edges of these materials, finding that across their range of interlayer spacings, all nanoscrolls have valence band edges lower than the oxygen evolution reaction (OER) potential of -5.67 eV, making them suitable as photoanodes. Furthermore, HfO\numsub{2}, PdO\numsub{2}, 1T-GeO\numsub{2}, and m-GeO\numsub{2} demonstrate conduction band edges that exceed the hydrogen evolution reaction potential of -4.44 eV, indicating their potential as single photocatalysts within specific interlayer spacing ranges.

To assess the hydrogen storage capabilities of the oxides, we performed $ab$ $initio$ molecular dynamics simulations of intercalated hydrogen species (H, H$^+$, and H\numsub{2}) within MoO\numsub{3} and 1T-GeO\numsub{2} nanoscrolls. Our results show that H atoms and H$^+$ ions adsorb onto the surface of the nanoscrolls, providing active sites for the HER. In contrast, H\numsub{2} molecules remain stable and are trapped within the scroll layers, enhancing the nanoscrolls' hydrogen storage capabilities utilizing their unique morphology. Remarkably, we found that these properties remain unchanged even in the presence of chalcogenide defects, due to the unique charge density arrangements on the oxide layers. Our research significantly expands the potential applications of oxide nanostructures in energy conversion and storage, paving the way for advancements in energy technologies.

\section{Acknowledgements}
This research was supported by the U.S. Department of Energy (DOE), Office of Science, Basic Energy Sciences (BES), under Award \# DE-SC0024184 (water-splitting study of heterostructures), by the National Science Foundation (NSF) under Award \# DMR-1906030 (studies of applications of nanoscrolls), and start-up funds from Arizona State University, USA. The authors also acknowledge the San Diego Supercomputer Center under the NSF-XSEDE Award No. DMR150006 and the Research Computing at Arizona State University for providing HPC resources. This research used resources of the National Energy Research Scientific Computing Center, a DOE Office of Science User Facility supported by the Office of Science of the U.S. Department of Energy under Contract No. DE-AC02-05CH11231. 

\section{Author Declarations}
The authors have no conflicts to disclose

\section{Supplementary Section}
Supplementary information contains tables detailing the electronic and structural properties of the selected materials, as well as calculation convergence parameters. It also includes figures of the QP band structures, electrostatic potentials for determination of band edges, and the schematics of structures. Finally, it includes further computational details for the determination of band edges.

\bibliography{references}

\end{document}